%% file: InfProdAmp.tex
\documentclass[11pt,a4paper]{article}
\pdfoutput=1

\usepackage{jheppub}

\usepackage{amsmath, amssymb, amsthm} 
\usepackage{mathtools}
\usepackage{thmtools}
\usepackage{bm}
\usepackage{dsfont}
\usepackage{braket}

\numberwithin{equation}{section}

\usepackage{enumitem}
\setlist[itemize]{noitemsep}
\setlist[itemize]{leftmargin=*}
\setlist[description]{noitemsep}

\usepackage{xcolor}
\usepackage{tikz}
\usepackage{pgfplots}
\usetikzlibrary{intersections,backgrounds}
\pgfplotsset{compat = newest}

\usepackage[numbers, sort&compress]{natbib}
\usepackage{hyperref}
\hypersetup{colorlinks=true, linktoc=page, linkcolor=purple, citecolor=blue}

\input{preambledefs}

\begin{document}

\title{Properties of infinite product amplitudes: Veneziano, Virasoro, and Coon}
    
\date{\today}

\author[a]{Nicholas Geiser}
\author[a]{and Lukas W. Lindwasser}

\affiliation[a]{
    Mani L. Bhaumik Institute for Theoretical Physics\\
    Department of Physics and Astronomy\\
    University of California, Los Angeles, CA 90095, USA}

\emailAdd{ngeiser@physics.ucla.edu}
\emailAdd{lukaslindwasser@physics.ucla.edu}


\abstract{
We detail the properties of the Veneziano, Virasoro, and Coon amplitudes. These tree-level four-point scattering amplitudes may be written as infinite products with an infinite sequence of simple poles. Our approach for the Coon amplitude uses the mathematical theory of $q$-analysis. We interpret the Coon amplitude as a $q$-deformation of the Veneziano amplitude for all $q \geq 0$ and discover a new transcendental structure in its low-energy expansion. We show that there is no analogous $q$-deformation of the Virasoro amplitude.}


\maketitle

\newpage

\section{Introduction}
\label{sec:intro}

In this paper, we shall detail the properties of the Veneziano~\cite{Veneziano:1968yb}, Virasoro~\cite{Virasoro:1969me}, and Coon~\cite{Coon:1969yw} amplitudes with zero Regge intercept.\footnote{For simplicity we only consider the scattering of four massless bosonic states. In this case, the tree-level open and closed superstring amplitudes respectively reduce to the Veneziano and Virasoro amplitudes with zero intercept.} These amplitudes describe the tree-level scattering of four massless particles and may be written as infinite products with an infinite sequence of simple poles. For each amplitude, we shall discuss its unitarity, high-energy behavior, low-energy expansion, and number theoretic properties. We shall synthesize these properties in a unified manner to facilitate comparison between the amplitudes.

\sm

Two of these amplitudes are well-known. The Veneziano amplitude describes the scattering of four open strings. The Virasoro amplitude describes the scattering of four closed strings. The lesser-known Coon amplitude is a one-parameter deformation of the Veneziano amplitude with a real deformation parameter $q \geq 0$. At $q=0$, the Coon amplitude reduces to a field theory amplitude. At $q=1$, the Coon amplitude is equal to the Veneziano amplitude.

\sm

Our approach for the Coon amplitude uses the mathematical theory of $q$-deformations, or $q$-analysis. Using a well-established $q$-deformation of the gamma function, we write a new single formula~\eqref{eq:CoonqGam} for the Coon amplitude valid for all $q \geq 0$. Previously, the Coon amplitude with~$q<1$ and the Coon amplitude with~$q>1$ were considered as distinct~\cite{Baker:1976en}. Our calculations confirm and extend the recent analysis of the Coon amplitude with~$q<1$~\cite{Figueroa:2022onw} to all $q \geq 0$. Moreover, we compute a compact formula~\eqref{eq:CoonLE} for the low-energy expansion of the Coon amplitude and discover a novel transcendental structure analogous to the number theoretic structure of the low-energy expansions of the Veneziano and Virasoro amplitudes.

\sm

As a function of $q$, the Coon amplitude demonstrates a subtle interplay between the properties of unitarity and meromorphicity (in the Mandelstam variables). For~${0<q<1}$, the Coon amplitude is unitary and non-meromorphic with an accumulation point spectrum.\footnote{A theory has an accumulation point spectrum if for some finite $M^2 > 0$, the number of particles with mass $m^2 < M^2$ is infinite.} For~${q>1}$, the Coon amplitude is non-unitary and meromorphic with no accumulation point. Only the Veneziano amplitude at $q=1$ is unitary and meromorphic with no accumulation point.

\sm

In any case, the Coon amplitude is a fruitful example for the study of general scattering amplitudes~\cite{Caron-Huot:2016icg, Huang:2020nqy, Maity:2021obe, Figueroa:2022onw, Huang:2022mdb}. While there is yet no definitive field theory or string worldsheet realization of the Coon amplitude, accumulation point spectra like those exhibited by the Coon amplitude with $q<1$ have been recently found in a stringy setup involving open strings ending on a D-brane~\cite{Maldacena:2022ckr}. Similar accumulation point spectra have also appeared in recent amplitude studies~\cite{Caron-Huot:2016icg, Ridkokasha:2020epy, Huang:2020nqy, Bern:2021ppb, Huang:2022mdb, Maldacena:2022ckr}. Famously, the hydrogen atom has energy levels $E_n = -13.6 \text{ eV} / n^2$ with an accumulation at $E_\infty = 0$.

\sm

In both this paper and in a companion paper~\cite{NGLL:2022}, we show that there is no naive $q$-deformation of the Virasoro amplitude or Virasoro-Coon amplitude analogous to our interpretation of the Coon amplitude as a $q$-deformed Veneziano amplitude. In this paper, we attempt and fail to construct a $q$-deformed Virasoro amplitude using functions from $q$-analysis. Our assumptions include crossing symmetry and polynomial residues on the same sequence of poles as the Coon amplitude. 

\sm

In~\cite{NGLL:2022}, we revisit this question by analyzing so-called generalized Veneziano and generalized Virasoro amplitudes, which are defined by generalizing the infinite product representations of the Veneziano and Virasoro amplitudes, respectively. Our procedure is an extension and clarification of Coon's original argument~\cite{Coon:1969yw} and related work~\cite{Fairlie:1994ad} in which we search for tree-level amplitudes with an infinite product form. We again assume crossing symmetry, but we now demand physical residues on an a priori unspecified sequence of poles~$ \lambda_n $. In other words, we do not assume the mass spectrum (while we do assume the mass spectrum in our search for a $q$-deformed Virasoro amplitude here). Under these assumptions, we find that the poles~$ \lambda_n $ must satisfy an over-determined set of non-linear recursion relations. The recursion relations for the generalized Veneziano amplitudes can be solved analytically. In the generalized Virasoro case, we numerically demonstrate that the only consistent solution to these recursion relations is the string spectrum. The Veneziano, Virasoro, and Coon amplitudes detailed here are in fact the healthiest amplitudes we find in~\cite{NGLL:2022}. This conclusion strengthens our present findings. 

\sm

There may also be a simple physical argument for the non-existence of a Virasoro analog of the Coon amplitude. We recall that accumulation point spectra were recently found in a stringy setup involving open strings ending on a D-brane~\cite{Maldacena:2022ckr}. If accumulation point spectra in string theory generically require open strings, then there may be no consistent $q$-deformed Virasoro amplitude with an accumulation point for $q<1$ since the Virasoro amplitude at $q=1$ describes the scattering of closed strings. These ideas are speculative and worthy of future research.

\sm

Our approach is part of the modern S-matrix bootstrap program, which attempts to construct general amplitudes which satisfy various physical properties such as unitarity, crossing, and analyticity without relying on an underlying dynamical theory~\cite{Correia:2020xtr}. The modern S-matrix bootstrap is a revival of an old approach~\cite{Eden:1966dnq} which predates modern quantum field theory (QFT) and attempts to constrain the space of physical theories, including those which may not be describable by QFT.

\subsection{Conventions}

We only consider tree-level scattering amplitudes for four massless particles in $d \geq 3$ spacetime dimensions. Such amplitudes have simple poles only. In a unitary theory, the residues of these poles equal finite sums of Gegenbauer polynomials,
\begin{align}
    C_j^{(\frac{d-3}{2})}(\cos\theta)
\end{align}
with positive coefficients. The positivity of these coefficients encodes the unitarity of the theory. The Mandelstam variables for this process are,
\begin{align}
    s
    &=
    -(p_1+p_2)^2
    =
    \phantom{-} 4 E^2
    \phantom{(1-\cos\theta)}
    \geq 0
\no \\
    t
    &=
    -(p_1+p_4)^2
    =
    -2 E^2 (1-\cos\theta)
    \leq 0
\no \\
    u
    &=
    -(p_1+p_3)^2
    = -2 E^2 (1+\cos\theta)
    \leq 0
\end{align}
and satisfy the mass-shell relation $s+t+u=0$. Here $E$ and $\theta$ are the center-of-mass energy and scattering angle, respectively, and the inequalities refer to the physical scattering regime. Since $s$-channel and $t$-channel Feynman diagrams correspond to the same cyclic ordering, color-ordered amplitudes (e.g. gluon amplitudes) will have only $s$-channel and $t$-channel poles and shall be denoted by $\cA(s,t)$. Amplitudes with poles in all three channels (e.g. graviton amplitudes) shall be denoted by $\cA(s,t,u)$. We use units in which the lowest massive state of any particular theory has mass~${m^2 = 1}$. In open (closed) string theory, this choice corresponds to~${\alpha' = 1}$ (${\alpha' = 4}$).

\subsection{Outline}

In \autoref{sec:Ven}, \autoref{sec:Vir}, and \autoref{sec:Coon} we discuss the properties of the Veneziano, Virasoro, and Coon amplitudes, respectively. In~\autoref{sec:CoonVir}, we attempt and fail to construct a Virasoro-Coon amplitude by $q$-deforming the Virasoro amplitude. The appendices contain various technical details. In~\autoref{sec:Gamma}, we review some properties of the gamma function. In~\autoref{sec:Geg}, we review some properties of the Gegenbauer polynomials. In~\autoref{sec:CoonLE}, we derive the low-energy expansion of the Coon amplitude.

\subsection*{Acknowledgements}

We are grateful to (in alphabetical order) Maor Ben-Shahar, Eric D'Hoker, Enrico Herrmann, Callum Jones, Dimitrios Kosmopoulos, Per Kraus, Oliver Schlotterer, and Terry Tomboulis for various discussions related to this work. NG is supported by a National Science Foundation (NSF) grant supplement from the Alliances for Graduate Education and the Professoriate Graduate Research Supplements (AGEP-GRS). NG and LL are supported by the Mani L. Bhaumik Institute for Theoretical Physics.


\section{The Veneziano amplitude}
\label{sec:Ven}

The Veneziano amplitude was discovered in 1968~\cite{Veneziano:1968yb} and describes the scattering of four open strings. More recently, the Veneziano amplitude has been revisited in the context of the modern S-matrix bootstrap program~\cite{Caron-Huot:2016icg, Green:2019tpt, Huang:2020nqy, Arkani-Hamed:2020blm, Maity:2021obe, Arkani-Hamed:2022gsa}.

\sm

In $d \leq 10$ dimensions, the Veneziano amplitude is a physically-admissible UV-completion of the tree-level four-point amplitude of maximally supersymmetric Yang-Mills field theory. The color-stripped tree-level field theory amplitude which describes the scattering of any four massless particles in the Yang–Mills supermultiplet is given by,
\begin{align}
\label{eq:SYM}
    \cA_{\text{SYM}}
    &=
    P_4 \, \frac{1}{st}
\end{align}
where $P_4$ is a kinematic prefactor which is determined by maximal supersymmetry and which contains the information about the particular states being scattered. For the four-gluon amplitude, schematically $P_4 = F^4$ where $F$ is the linearized field strength. The second factor $\tfrac{1}{st}$ is symmetric in $(s,t)$ and is a meromorphic function with simple poles from massless particle exchange in the $s$-channel and $t$-channel. In the high-energy Regge limit $s \to \infty$ with fixed polarizations and fixed $t < 0$, the prefactor $P_4 \propto s^2$, and the amplitude diverges as $\cA_{\text{SYM}} \propto s$.

\sm

In tree-level open superstring theory, the color-stripped amplitude which describes the scattering process~\eqref{eq:SYM} is given by,
\begin{align}
\label{eq:open}
    \cA_{\text{open}}
    &=
    P_4 \, \cA_{\text{Ven}}(s,t)
\end{align}
where $\cA_{\text{Ven}}$ is the Veneziano amplitude,
\begin{align}
\label{eq:Ven}
    \cA_{\text{Ven}}(s,t)
    &=
    \frac{ \Gamma(-s) \Gamma(-t) }{ \Gamma(1-s-t) }
\end{align}
Like the corresponding field theory factor, the Veneziano amplitude is symmetric in~$(s,t)$ and is a meromorphic function with simple poles only. We may explicitly exhibit these poles using the infinite product representation~\eqref{eq:GammaIP} of the gamma function to write,
\begin{align}
\label{eq:VenIP}
    \cA_{\text{Ven}}(s,t)
    &=
    \frac{1}{st \mathstrut}
    \prod_{n \geq 1}
    \frac{ n^2 - n (s+t) }{ (s-n) (t-n) }
\end{align}
The infinite sequence of massive poles at $s = 1, 2, \dots$ correspond to excited stringy states.

\subsection{Unitarity}

The kinematic prefactor $P_4$ which appears in both the field theory amplitude~\eqref{eq:SYM} and the open superstring amplitude~\eqref{eq:open} has a positive expansion on the Gegenbauer polynomials in $d \leq 10$ dimensions~\cite{Arkani-Hamed:2022gsa, Huang:2022mdb}. It remains then to check the unitarity of the Veneziano amplitude itself. On the massless $s$-channel pole,~$P_4 \propto t^2$, and the residue of the Veneziano amplitude agrees with field theory,
\begin{align}
    \Res_{s=0} \cA_{\text{Ven}}(s,t)
    &=
    \Res_{s=0} \frac{1}{st}
    =
    \frac{1}{t}
    \quad
    \implies
    \quad
    \Res_{s=0} \cA_{\text{open}}
    \propto t
\intertext{indicating the exchange of a massless spin-$1$ state, the gluon. The residue of each massive pole at $s=N \geq 1$ is a degree-$(N-1)$ polynomial in $t$,}
\label{eq:VenRes}
    \Res_{s=N}
    \cA_{\text{Ven}}(s,t)
    &=
    \frac{1}{N!}
    (t+1)(t+2) \cdots (t+N-1)
\intertext{indicating the exchange of states with mass $m^2 = N$ and spins $j \leq N+1$. These residues may be expanded in terms of Gegenbauer polynomials using the identities in~\autoref{sec:Geg},}
    \Res_{s=N}
    \cA_{\text{Ven}}(s,t)
    &=
    \sum_{j=0}^{N-1}
    c^{\mathstrut}_{N,j} \,
    C^{(\frac{d-3}{2})}_j
    \big( 1+\tfrac{2t}{N} \big)
\end{align}
with the first few coefficients given by,
\begin{align}
    c_{1,0}
    &=
    1
    &
    c_{2,0}
    &=
    0
    &
    c_{3,0}
    &=
    \tfrac{10-d\mathstrut}{24(d-1)}
\no \\
    &&
    c_{2,1}
    &=
    \tfrac{1\mathstrut}{2(d-3)}
    &
    c_{3,1}
    &=
    0
\no \\
    &&
    &&
    c_{3,2}
    &=
    \tfrac{3\mathstrut}{4(d-1)(d-3)}
\end{align}
The coefficient $c_{3,0}$ is negative for $d > 10$, indicating the non-unitarity of the superstring above its critical dimension $d=10$. These coefficients were recently studied in $d=4$~\cite{Maity:2021obe} and were recently shown to be positive for all $d \leq 6$~\cite{Arkani-Hamed:2022gsa}. The unitarity of superstring theory in $d \leq 10$ (and thus the positivity of the $c_{N,j}$) is known from the no-ghost theorem~\cite{Brower:1972wj, Goddard:1972iy, Thorn:1983cz}, but there is yet no direct proof that $c_{N,j} > 0$ for all $d \leq 10$.

\subsection{High-energy}

The high-energy behavior of the Veneziano amplitude may be calculated using Stirling's formula~\eqref{eq:Stirling}. In the Regge limit of large $|s| \gg 1$ with phase ${0 < \arg(s) < 2\pi}$ (to avoid the poles of the gamma function) and fixed $t < 0$, we find,
\begin{align}
\label{eq:VenReg}
    \cA_{\text{Ven}}(s,t)
    \widesim[2]{|s| \to \infty}
    (-s)^{t-1} \,
    \Gamma(-t) \,
    \big(
    1 + \cO( s^{-1} )
    \big)
\end{align}
Compared to field theory, the extra exponent $t$ softens the UV behavior. For any scattered states with fixed polarizations, there is a range of fixed $t < 0$ such that ${\lim_{|s| \to \infty} \cA_{\text{open}} = 0}$ while this limit diverges in the corresponding field theory amplitude.

\subsection{Low-energy}

At leading order in the low-energy expansion $|s|, |t| \ll 1$, the Veneziano amplitude reproduces field theory. At higher order, stringy corrections to field theory are given in terms of Riemann zeta-values. Using the Taylor expansion for the gamma function~\eqref{eq:GammaTay}, we find,
\begin{align}
\label{eq:VenLE}
    \cA_{\text{Ven}}(s,t)
    &=
    \frac{1}{st} \,
    \exp \,
    \sum_{k \geq 2} \,
    \frac{\zeta(k)}{k}
    \big[ s^k + t^k - (s+t)^k \big]
\no \\
    &=
    \frac{1}{st}
    - \zeta(2)
    - \zeta(3) \, (s+t)
    - \zeta(4) \, ( s^2 + \tfrac{1}{4} st + t^2 )
    + \cdots
\end{align}
The Veneziano amplitude exhibits a remarkable property called uniform transcendentality, meaning each term in its low-energy expansion may be assigned the same transcendental weight. If we assign weight~$k$ to the zeta-value~$\zeta(k)$ (the standard number theoretic assignment) and weight~$-1$ to the Mandelstam variables, then each term in~\eqref{eq:VenLE} has transcendental weight two. Uniform transcendentality is in fact a general property of tree-level superstring amplitudes~\cite{Schlotterer:2012ny}, and the transcendental structure of one-loop superstring amplitudes is under active study~\cite{DHoker:2019blr, DHoker:2021ous}. In comparison, non-trivial transcendental structure in field theory only arises from loop integrals~\cite{Kotikov:2002ab, Beccaria:2009vt, Arkani-Hamed:2010pyv, Broedel:2018qkq}.


\section{The Virasoro amplitude}
\label{sec:Vir}

The Virasoro amplitude was discovered in 1969~\cite{Virasoro:1969me} and describes the scattering of four closed strings. Like the Veneziano amplitude, the Virasoro amplitude has also been recently revisited in the context of the modern S-matrix bootstrap program~\cite{Green:2019tpt, Arkani-Hamed:2020blm, Arkani-Hamed:2022gsa}.

\sm

In $d \leq 10$ dimensions, the Virasoro amplitude is a physically-admissible UV-completion of the tree-level four-point amplitude of maximal supergravity. The tree-level field theory amplitude which describes the scattering of any four massless particles in the supergravity multiplet is given by,
\begin{align}
\label{eq:SUGRA}
    \cA_{\text{SG}}
    &=
    P_8 \Big( {-\frac{1}{stu}} \Big)
\end{align}
where $P_8$ is a kinematic prefactor which is determined by maximal supersymmetry and which contains the information about the particular states being scattered. For the four-graviton amplitude, schematically $P_8 = R^4$ where $R$ is the linearized Riemann curvature. The second factor $-\tfrac{1}{stu}$ is symmetric in $(s,t,u)$ and contains poles from massless particle exchange in the $s$-channel, $t$-channel, and $u$-channel. In the high-energy Regge limit $s \to \infty$ with fixed polarizations and fixed $t < 0$, the prefactor $P_8 \propto s^4$, and the amplitude diverges as $ \cA_{\text{SG}} \propto s^2$.

\sm

In tree-level closed superstring theory, the amplitude which describes the scattering process~\eqref{eq:SUGRA} is given by,
\begin{align}
\label{eq:closed}
    \cA_{\text{closed}}
    &=
    P_8 \, \cA_{\text{Vir}}(s,t,u)
\end{align}
where $\cA_{\text{Vir}}$ is the Virasoro amplitude,
\begin{align}
\label{eq:Vir}
    \cA_{\text{Vir}}(s,t,u)
    &=
    \frac{ \Gamma(-s) \Gamma(-t) \Gamma(-u) }
        { \Gamma(1+s) \Gamma(1+t) \Gamma(1+u) }
\end{align}
Like the corresponding field theory factor, the Virasoro amplitude is symmetric in~$(s,t,u)$ and is a meromorphic function with simple poles only. We may explicitly exhibit these poles using the infinite product representation~\eqref{eq:GammaIP} of the gamma function to write,
\begin{align}
\label{eq:VirIP}
    \cA_{\text{Vir}}(s,t,u)
    &=
    - \frac{1}{stu \mathstrut}
    \prod_{n \geq 1}
    \frac{ -n^3 - n (st+tu+us) - stu }{ (s-n)(t-n)(u-n) }
\end{align}
The infinite sequence of massive poles at $s = 1, 2, \dots$ correspond to excited stringy states.

\subsection{Unitarity}

The kinematic prefactor $P_8$ which appears in both the field theory amplitude~\eqref{eq:SUGRA} and the closed superstring amplitude~\eqref{eq:closed} has a positive expansion on the Gegenbauer polynomials in $d \leq 10$ dimensions~\cite{Arkani-Hamed:2022gsa, Huang:2022mdb}. It remains then to check the unitarity of the Virasoro amplitude itself. On the massless $s$-channel pole,~$P_8 \propto t^4$, and the residue of the Virasoro amplitude agrees with field theory,
\begin{align}
    \Res_{s=0} \cA_{\text{Vir}}(s,t,-s-t)
    &=
    \Res_{s=0} \frac{1}{st(s+t)}
    =
    \frac{1}{t^2}
    \quad
    \implies
    \quad
    \Res_{s=0} \cA_{\text{closed}}
    \propto t^2
\intertext{indicating the exchange of a massless spin-$2$ state, the graviton. The residue of each massive pole at $s=N \geq 1$ is a degree-$(2N-2)$ polynomial in $t$, indicating the exchange of states with mass $m^2 = N$ and spins $j \leq 2N+2$. In fact, the residues of the Virasoro amplitude equal the residues of the Veneziano amplitude~\eqref{eq:VenRes} squared,}
\label{eq:VirRes}
    \Res_{s=N}
    \cA_{\text{Vir}}(s,t,-s-t)
    &=
    \Big\{
    \frac{1}{N!}
    (t+1)(t+2) \cdots (t+N-1)
    \Big\}^2
\no \\
    &=
    \Big\{
    \Res_{s=N}
    \cA_{\text{Ven}}(s,t)
    \Big\}^2
\intertext{These residues may be expanded in terms of Gegenbauer polynomials using the identities in~\autoref{sec:Geg},}
    \Res_{s=N}
    \cA_{\text{Vir}}(s,t,-s-t)
    &=
    \sum_{j=0}^{2N-2}
    c^{\mathstrut}_{N,j} \,
    C^{(\frac{d-3}{2})}_j
    \big( 1+\tfrac{2t}{N} \big)
\end{align}
with the first few coefficients given by,
\begin{align}
    c_{1,0}
    &=
    1
    &
    c_{2,0}
    &=
    \tfrac{1 \mathstrut}{4(d-1)}
    &
    c_{3,0}
    &=
    \tfrac{224-18d+d^2 \mathstrut}{576(d+1)(d-1)}
\no \\
    &&
    c_{2,1}
    &=
    0
    &
    c_{3,1}
    &=
    0
    \vphantom{\tfrac{\mathstrut}{\mathstrut}}
\no \\
    &&
    c_{2,2}
    &=
    \tfrac{1 \mathstrut}{2(d-1)(d-3)}
    &
    c_{3,2}
    &=
    \tfrac{24-d \mathstrut}{16(d+3)(d-1)(d-3)}
\no \\
    &&
    &&
    c_{3,3}
    &=
    0
    \vphantom{\tfrac{\mathstrut}{\mathstrut}}
\no \\
    &&
    &&
    c_{3,4}
    &=
    \tfrac{27 \mathstrut}{8(d+3)(d+1)(d-1)(d-3)}
\end{align}
The positivity of these coefficients below the critical dimension follows indirectly from the no-ghost theorem~\cite{Brower:1972wj, Goddard:1972iy, Thorn:1983cz}, but there is yet no direct proof that $c_{N,j} > 0$ for all $d \leq 10$.

\subsection{High-energy}

The high-energy behavior of the Virasoro amplitude may be calculated using Stirling's formula~\eqref{eq:Stirling}. In the Regge limit of large $|s| \gg 1$ with phase ${0 < \arg(s) < \pi}$ (to avoid the poles of the gamma function) and fixed $t < 0$, we find,
\begin{align}
\label{eq:VirReg}
    \cA_{\text{Vir}}(s,t,-s-t)
    \widesim[2]{|s| \to \infty}
    (-s)^{t-1} \,
    s^{t-1} \,
    \frac{ \Gamma(-t) }{ \Gamma(1+t) } \,
    \big(
    1 + \cO( s^{-1} )
    \big)
\end{align}
Compared to field theory, the extra exponent $2t$ softens the UV behavior. For any scattered states with fixed polarizations, there is a range of fixed $t < 0$ such that ${\lim_{|s| \to \infty} \cA_{\text{closed}} = 0}$ while this limit diverges in the corresponding field theory amplitude.

\subsection{Low-energy}

At leading order in the low-energy expansion $|s|, |t|, |u| \ll 1$, the Virasoro amplitude reproduces field theory. At higher order, the stringy corrections to field theory are given in terms of Riemann zeta-values. Using the Taylor expansion for the gamma function~\eqref{eq:GammaTay}, we find,
\begin{align}
\label{eq:VirLE}
    \cA_{\text{Vir}}(s,t,u)
    &=
    - \frac{1}{stu} \,
    \exp \,
    \sum_{k \geq 1} \,
    \frac{2 \, \zeta(2k+1)}{2k+1}
    ( s^{2k+1} + t^{2k+1} + u^{2k+1} )
\no \\
    &=
    - \frac{1}{stu}
    - 2 \, \zeta(3)
    - \zeta(5) \, ( s^2 + t^2 + u^2 )
    + \cdots
\end{align}
Like the Veneziano amplitude~\eqref{eq:VenLE}, the low-energy expansion of the Virasoro amplitude exhibits uniform transcendentality. If we assign weight~$k$ to~$\zeta(k)$ and weight~$-1$ to the Mandelstam variables, then each term in~\eqref{eq:VirLE} has transcendental weight three.

\sm

Furthermore, we note that only odd zeta-values occur in~\eqref{eq:VirLE} while both even and odd zeta-values occurred in the low-energy expansion of the Veneziano amplitude~\eqref{eq:VenLE}. This discrepancy between the Veneziano and Virasoro amplitudes may be described by the so-called single-valued map, which maps the (motivic) zeta-values $\zeta(k)$ to the single-valued zeta-values $\zeta_{\rm{sv}}(k)$, defined by,
\begin{align}
    \zeta_{\rm{sv}}(2k)
    &=
    0
    &
    \zeta_{\rm{sv}}(2k+1)
    &=
    2 \, \zeta(2k+1)
\end{align}
The single-valued zeta-values are so-called because they descend from single-valued versions of the multi-valued polylogarithm functions $\operatorname{Li}_k(z)$, which evaluate to the Riemann zeta function at $z=1$,
\begin{align}
    \operatorname{Li}_{k}(z)
    =
    \sum_{n \geq 1}
    \frac{ z^n }{ n^k }
    \quad
    \xrightarrow[z \to 1]{}
    \quad
    \zeta(k)
    =
    \sum_{n \geq 1}
    \frac{ 1 }{ n^k }
\end{align}
Comparing~\eqref{eq:VenLE} and~\eqref{eq:VirLE}, we see that the Veneziano amplitude becomes the Virasoro amplitude under the single-valued map acting term-by-term on the low-energy expansion,
\begin{align}
\label{eq:sv}
    (st) \, \cA_{\text{Ven}}
    \xrightarrow[\rm{sv}]{}
    (-stu) \, \cA_{\text{Vir}}
\end{align}
General tree-level open and closed superstring amplitudes are in fact related by the single-valued map~\cite{Schlotterer:2012ny, Brown:2013gia, Stieberger:2013wea, Stieberger:2014hba, Brown:2018omk, Schlotterer:2018zce, Vanhove:2018elu, Brown:2019wna}, encoding a deep number theoretical relationship between the open and closed superstrings, and thus between gauge theories and theories of gravity.

\sm

Both the single-valued map~\eqref{eq:sv} and the residue relation~\eqref{eq:VirRes} are manifestations of another relationship between open and closed superstring amplitudes. The Kawai-Lewellen-Tye (KLT) relations~\cite{Kawai:1985xq} express tree-level closed superstring amplitudes as bilinears of tree-level open superstring amplitudes. Informally, the closed superstring is equal to the open superstring squared. The four-point KLT relation is,
\begin{align}
    \cA_{\text{Vir}}(s,t,u)
    =
    \cA_{\text{Ven}}(s,t)
    \,
    S_{\text{KLT}}
    \,
    \cA_{\text{Ven}}(s,t)
\end{align}
where $S_{\text{KLT}}$ is the KLT kernel,
\begin{align}
    S_{\text{KLT}}
    =
    \frac{ \sin( \pi s) \sin( \pi t) }
         { \pi \sin \big( \pi (s+t) \big) }
\end{align}
This expression for $S_{\text{KLT}}$ follows from the definition of the Veneziano amplitude~\eqref{eq:Ven}, the definition of the Virasoro amplitude~\eqref{eq:Vir}, and the reflection formula for the gamma function~\eqref{eq:GammaRefl}. In the field theory (low-energy) limit, the KLT relations are known as the double copy between gauge theory and gravity~\cite{Bern:2010ue, Bern:2019prr}.


\section{The Coon amplitude}
\label{sec:Coon}

The Coon amplitude was discovered in 1969 as a generalization of the Veneziano amplitude with non-linear Regge trajectories~\cite{Coon:1969yw}. The subsequent studies of the Coon amplitude were phenomenologically motivated. A concise survey of this early literature is given in a (quite difficult to find) 1989 review~\cite{Romans:1989di} (which cites an expanded but unpublished pre-print~\cite{Romans:1988qs} which we could not locate). Around this time, the Coon amplitude was revisited in the broader context of string theory~\cite{Romans:1989di, Fairlie:1994ad}. Most recently, the Coon amplitude has reappeared in the modern S-matrix bootstrap program~\cite{Caron-Huot:2016icg, Ridkokasha:2020epy, Huang:2020nqy, Maity:2021obe, Figueroa:2022onw, Huang:2022mdb, Maldacena:2022ckr}.

\sm

The Coon amplitude $\cA_q$ is a one-parameter deformation of the Veneziano amplitude with a real deformation parameter $q \geq 0$. To construct a full four-point scattering amplitude, we replace $\cA_{\text{Ven}}$ in the open superstring amplitude~\eqref{eq:open} with $\cA_{q}$ to describe the scattering of four massless states in a putative $q$-deformed string theory,
\begin{align}
\label{eq:qstrings}
    \cA_{q\text{-strings}}
    &=
    P_4 \, \cA_q(s,t)
\end{align}
This deformation may be understood using the mathematical theory of $q$-deformations or $q$-analogs, also known as $q$-analysis. 

\subsection{\texorpdfstring{$q$}{q}-analysis}

In mathematics, a $q$-analog of a theorem, function, identity, or expression is a generalization involving a deformation parameter $q$ that returns the original mathematical object in the limit $q \to 1$. Many special functions and differential equations have well-studied $q$-analogs dating back to the nineteenth century~\cite{gasper_rahman_2004}. For our purposes, we shall only need a few $q$-ingredients. We first define the $q$-integers $[n]_q$ by,
\begin{align}
    [n]_q
    =
    \frac{1-q^n}{1-q^{\phantom{n}}}
    =
    1 + q + q^2 + \cdots + q^{n-1}
    \quad
    \xrightarrow[q \to 1]{}
    \quad
    n
\end{align}
where the second equality holds for $n \geq 1$.

\sm

In passing from the Veneziano amplitude~\eqref{eq:Ven} to the Coon amplitude, we shall replace the linear Regge trajectory $\alpha(s) = s$ that appears in the arguments of the gamma functions with a non-linear deformation $\alpha_q(s)$ that satisfies $\alpha_q([n]_q) = n$. The $q$-deformed Regge trajectory is thus,
\begin{align}
    \alpha_q(s)
    &=
    \frac{ \ln \big( 1+(q-1)s \big) }{ \ln q }
\end{align}
This Regge trajectory becomes linear as $\lim_{q \to 1} \alpha_q(s) = s$.

\sm

The gamma functions in the Veneziano amplitude~\eqref{eq:Ven} are themselves replaced by the so-called $q$-gamma function, which is defined for complex $q$ by~\cite{gasper_rahman_2004},
\begin{align}
\label{eq:qGam}
    \Gamma_q(z)
    &=
\begin{cases}
    \phantom{q^{\frac{z(z-1)}{2}}}
    (1-q)^{1-z}
    \displaystyle \prod_{n \geq 0}
    \frac{1-q^{+n+1}}{1-q^{+n+z}}
    \qquad
    &
    |q| < 1
\\[3ex]
    q^{\frac{z(z-1)}{2}}
    (q-1)^{1-z}
    \displaystyle \prod_{n \geq 0}
    \frac{1-q^{-n-1}}{1-q^{-n-z}}
    \qquad
    &
    |q| > 1
\end{cases}
\end{align}
The $q$-gamma function obeys a functional equation analogous to $\Gamma(z+1) = z \, \Gamma(z)$,
\begin{align}
    \Gamma_q(z+1)
    &=
    \frac{1-q^z}{1-q^{\phantom{z}}} \,
    \Gamma_q(z)
\end{align}
and becomes the ordinary gamma function as $\lim_{q \to 1^\pm} \Gamma_q(z) = \Gamma(z)$. Many properties of the gamma function have precise $q$-analogs. For instance, the asymptotic behavior of the $q$-gamma function is given by a $q$-analog of Stirling's formula~\cite{Moak1984},
\begin{align}
\label{eq:qStirling}
    \ln \Gamma_q(z)
    &\widesim[2]{|q^{z}| \to 0}
    (z-\half)
    \ln \frac{1-q^z}{1-q^{\phantom{z}}}
    + \frac{\operatorname{Li}_2 (1-q^z) }
           { \ln q }
    + \half \ln(2\pi)
    + C_{q}
    + \cO \big( q^z \big)
\end{align}
which is valid for small $| q^{z} | \ll 1$. Here $\operatorname{Li}_2(z)$ is the dilogarithm and $C_q$ is a known $q$-dependent constant.

\subsection{\texorpdfstring{$q$}{q}-deformed Veneziano}

In terms of these $q$-ingredients, the Coon amplitude for all $q \geq 0$ is given by,
\begin{align}
\label{eq:CoonqGam}
    \cA_{q}(s,t)
    &=
    q^{ \alpha_q(s) \alpha_q(t) - \alpha_q(s) - \alpha_q(t)} \,
    \frac{ \Gamma_q \big( {-\alpha_q(s)} \big)
           \Gamma_q \big( {-\alpha_q(t)} \big) }
         { \Gamma_q \big( 1-\alpha_q(s)-\alpha_q(t) \big) }
\end{align}
Our conventions for the Coon amplitude differ from the older literature by an overall normalization but are chosen so that its leading low-energy behavior is $\frac{1}{st} (1+\cO(s,t))$ to facilitate comparison to the Veneziano amplitude. Clearly, $\lim_{q \to 1} \cA_q = \cA_{\text{Ven}}$. Moreover, our single formula contains both the Coon amplitude with~$q<1$ and the Coon amplitude with~$q>1$, which were previously considered as distinct~\cite{Baker:1976en}. However, many properties of the Coon amplitude, including its meromorphicity as a function of the Mandelstam variables, are obscured in the form~\eqref{eq:CoonqGam}. The prefactor $\smash{q^{ \alpha_q(s) \alpha_q(t)}}$ is explicitly non-meromorphic, but we shall soon see that it is perfectly natural. 

\sm

Using the definition of $q$-gamma function, we may write the Coon amplitude in terms of one convergent infinite product for $q < 1$ and another for $q > 1$,
\begin{align}
\label{eq:CoonIP1}
    \cA_{q}(s,t)
    =
    q^{ \alpha_q(s) \alpha_q(t) } \,
    \Theta(1-q) \,
    &
    \frac{1}{st \mathstrut}
    \prod_{n \geq 1}
    \frac{ ( 1 - q^{-\alpha_q(s)-\alpha_q(t)+n} )
           ( 1 - q^{+n} ) }
         { ( 1 - q^{-\alpha_q(s)+n} )
           ( 1 - q^{-\alpha_q(t)+n} ) }
\no \\
    {}+{}
    \Theta(q-1) \,
    &
    \frac{1}{st \mathstrut}
    \prod_{n \geq 1}
    \frac{ ( 1 - q^{+\alpha_q(s)+\alpha_q(t)-n} )
           ( 1 - q^{-n} ) }
         { ( 1 - q^{+\alpha_q(s)-n} )
           ( 1 - q^{+\alpha_q(t)-n} ) }
\end{align}
where the step function is defined by $\Theta(x \geq 0)=1$ and $\Theta(x<0)=0$. This form is nice because the infinite product of each of the four factors in either infinite product separately converges. Moreover, for $q>1$, the non-meromorphic prefactor $\smash{q^{ \alpha_q(s) \alpha_q(t)}}$ has canceled against similar non-meromorphic factors in the $q$-gamma functions.

\sm

We may further massage~\eqref{eq:CoonIP1} into a form with just one infinite product for all~$q \geq 0$ times a piecewise $q$-dependent prefactor,
\begin{align}
\label{eq:CoonIP2}
    \cA_{q}(s,t)
    &=
    \bigg\{
    q^{ \frac{ \ln ( 1+(q-1)s ) }{ \ln q }
        \frac{ \ln ( 1+(q-1)t ) }{ \ln q } }
    \, \Theta(1-q)
    + \Theta(q-1)
    \bigg\}
\no \\
    & \quad
    \times
    \frac{1}{st \mathstrut}
    \prod_{n \geq 1}
    \frac{ \big( \frac{1-q^n}{1-q^{\phantom{n}}} \big)^2
            - \big( \frac{1-q^n}{1-q^{\phantom{n}}} \big) (s+t)
            + (1-q^n) st }
        { \big( s - \frac{1-q^n}{1-q^{\phantom{n}}} \big)
          \big( t - \frac{1-q^n}{1-q^{\phantom{n}}} \big) }
\end{align}
We must take care with this expression because the infinite products of the numerator and denominator do not separately converge. In this form, we see that the Coon amplitude with~${q > 0}$ has an infinite sequence of simple poles in both the $s$-channel and $t$-channel. These poles occur at the $q$-integers,
\begin{align}
    [n]_q
    =
    \frac{1-q^n}{1-q^{\phantom{n}}}
\end{align}
for integer $n \geq 0$. For $q=0$, the infinite product is just $\prod_{n \geq 1} 1 = 1$. For $0 < q < 1$, the poles tend to an accumulation point at $\frac{1}{1-q}$. For~$q \geq 1$, the poles tend to infinity, and $q=1$ reproduces the string spectrum.

\sm

Like the Veneziano amplitude, the Coon amplitude is symmetric in~$(s,t)$ and has simple poles only. For $q \geq 1$, the Coon amplitude is meromorphic, but for $q<1$, the Coon amplitude as written in~\eqref{eq:CoonIP2} has an explicit non-meromorphic factor $q^{ \alpha_q(s) \alpha_q(t) }$ with branch cuts in the complex $s$-plane and $t$-plane starting at the accumulation point $\frac{1}{q-1}$. In the limits $q \to 0$ and $q \to 1^-$, this non-meromorphic prefactor becomes $q^{ \alpha_q(s) \alpha_q(t) } \to 1$, and the Coon amplitude reproduces the field theory and Veneziano amplitudes, respectively,
\begin{align}
    \cA_{q}(s,t)
    \quad
    & \xrightarrow[q \to 0]{}
    \quad
    \mask{\cA_{\text{Ven}}(s,t)}{\frac{1}{st}}
\no \\
    \cA_{q}(s,t)
    \quad
    & \xrightarrow[q \to 1]{}
    \quad
    \cA_{\text{Ven}}(s,t)
\end{align}

\subsection{Unitarity}

To compute the residues of the Coon amplitude, we shall manipulate the form~\eqref{eq:CoonIP1} with manifestly convergent infinite products. The algebra is tedious but straightforward. The residue of the massless pole agrees with field theory for all $q \geq 0$,
\begin{align}
    \Res_{s=0} \cA_{q}(s,t)
    &=
    \Res_{s=0} \frac{1}{st}
    =
    \frac{1}{t}
\intertext{For $q > 0$, the residue of the massive pole at $s=[N]_q$ with $N \geq 1$ is a degree-$(N-1)$ polynomial in $t$, indicating the exchange of states with mass $m^2 = [N]_q$ and spins $j \leq N+1$,}
\label{eq:CoonRes}
    \Res_{s=[N]_q}
    \cA_{q}(s,t)
    &=
    q^N \,
    \prod_{n=1}^{N}
    \frac{1}{ \big( \tfrac{1-q^n}{1-q^{\phantom{n}}} \big) }
    \,
    \prod_{n=1}^{N-1}
    \big( q^n \, t + \tfrac{1-q^n}{1-q^{\phantom{n}}} \big)
\intertext{On the poles, the non-meromorphic factor $\smash{q^{ \alpha_q([N]_q) \alpha_q(t) } = ( 1+(q-1)t )^N}$ ensures that the residues are polynomials in $t$ for $q<1$. In any case, these residues may be expanded in terms of Gegenbauer polynomials using the identities in~\autoref{sec:Geg},}
    \Res_{s=[N]_q}
    \cA_{q}(s,t)
    &=
    \sum_{j=0}^{N-1}
    c^{\mathstrut}_{N,j} \,
    C^{(\frac{d-3}{2})}_j
    \big( 1+\tfrac{2t}{ [N]_q } \big)
\end{align}
with the first few coefficients given by,
\begin{align}
    c_{1,0}
    &=
    q
    \vphantom{\tfrac{\mathstrut}{\mathstrut}}
\no \\
    c_{2,0}
    &=
    \tfrac{q^2(1-q)(2+q)}{2(1+q)}
\no \\
    c_{2,1}
    &=
    \tfrac{q^3 \mathstrut}{2(d-3)}
\no \\
    c_{3,0}
    &=
    \tfrac{ q^3 [ 4(d-1)+2q(d-1)-6q^2(d-1)
        -q^3(5d-6)-2q^4(d-2)+3dq^5+2dq^6+dq^7 ]}
          { 4(d-1)(1+q)(1+q+q^2) }
\no \\
    c_{3,1}
    &=
    \tfrac{ q^4(1-q)(1+3q+2q^2+q^3) }
         { 2(d-3)(1+q) }
\no \\
    c_{3,2}
    &=
    \tfrac{ q^6(1+q+q^2) }
         { 2(d-1)(d-3)(1+q) }
\end{align}
For $q > 1$, the coefficient $c_{2,0}$ is negative in any number of dimensions, indicating non-unitarity. The non-unitarity of the Coon amplitude with $q>1$ has been known since the early 1970s~\cite{Baker:1971zs}. The unitarity of the Coon amplitude with $q<1$ is more subtle. This case was studied in the 1990s~\cite{Fairlie:1994ad} and again more recently~\cite{Figueroa:2022onw}. The most recent numerical studies indicate that the Coon amplitude with $q<1$ is unitary below some $q$-dependent critical dimension~\cite{Figueroa:2022onw}. This critical dimension is $d=10$ in the limit $q \to 1$ and $d = \infty$ in the limit $q \to 0$ .

\sm

Although the Coon amplitude with $q<1$ may be unitary, it is non-meromorphic due to the factor $\smash{q^{\alpha_q(s)\alpha_q(t)}}$. As we discussed above, this explicit non-meromorphic factor is necessary for the Coon amplitude to have polynomial residues. The Coon amplitude with~${q<1}$ also has an accumulation point of poles at $\frac{1}{1-q}$. By definition, meromorphic functions can only have isolated poles. Thus, the infinite product itself is non-meromorphic even without the explicit non-meromorphic factor.

\sm

For $q>1$ the situation is reversed. There the Coon amplitude is meromorphic with no accumulation point, but it is non-unitary. Only the Veneziano amplitude at $q=1$ is both meromorphic and unitary.

\subsection{High-energy}

The high-energy behavior of the Coon amplitude may be calculated using the $q$-analog of Stirling's formula~\eqref{eq:qStirling}. In the Regge limit with fixed $t < 0$ and large~${|s| \gg 1}$ with phase $0 < \arg(s) < 2\pi$ (to avoid the poles of the $q$-gamma function as well as the branch cut for~${0<q<1}$), we find,
\begin{align}
\label{eq:CoonReg}
    \cA_{q}(s,t)
    \widesim[2]{|s| \to \infty}
    (-s)^{\alpha_q(t)-1} \,
    \frac{ \Gamma_q \big( {-\alpha_q(t)} \big) }
         { (q-1)t+1 }
    \,
    \big[
    1 + \cO \big( (q-1)^{-1} s^{-1} \big)
    \big]
\end{align}
which agrees with the Regge limit~\eqref{eq:VenReg} of the Veneziano amplitude as ${q \to 1}$ (ignoring the subtlety that the small parameter blows up at $q=1$). For both $0<q<1$ and $q>1$, the exponent $\alpha_q(t) = \ln(1+(q-1)t) / \ln q$ can be made arbitrarily large and negative as a function of $t < 0$. For any scattered states with fixed polarizations, there is thus a range of fixed $t < 0$ such that ${\lim_{|s| \to \infty} \cA_{q\text{-strings}} = 0}$ while this limit diverges in the corresponding field theory amplitude.

\subsection{Low-energy}

The low-energy expansion of the Coon amplitude with $q<1$ was recently studied~\cite{Figueroa:2022onw}. Here we extend that result to all $q \geq 0$. The details of our calculation are given in~\autoref{sec:CoonLE}.

\sm

Like the Veneziano amplitude, the Coon amplitude reproduces field theory at leading order. At higher order and for all~$q > 0$, corrections to field theory are given in terms of the $q$-deformation $\operatorname{Li}_k(z;q)$ of the polylogarithm $\operatorname{Li}_k(z)$, which evaluates to the Riemann zeta function $\zeta(k)$ at $q=z=1$~\cite{Schlesinger:2001},
\begin{align}
    \operatorname{Li}_{k}(z;q)
    =
    \sum_{n \geq 1}
    \frac{ z^n }
         { \big( \frac{1-q^n}{1-q^{\phantom{n}}} \big)^{k} }
    \quad
    \xrightarrow[q \to 1]{}
    \quad
    \operatorname{Li}_{k}(z)
    =
    \sum_{n \geq 1}
    \frac{ z^n }{ n^k }
    \quad
    \xrightarrow[z \to 1]{}
    \quad
    \zeta(k)
    =
    \sum_{n \geq 1}
    \frac{ 1 }{ n^k }
\end{align}
The low-energy expansion includes in particular the $q$-deformed polylogarithms $\operatorname{Li}_{k}(q^j;q)$ with integers $k > j \geq 1$. For all $q \geq 0$, the defining sums for these special functions are absolutely convergent and finite,
\begin{align}
    \operatorname{Li}_{k}(q^j;q)
    =
    \sum_{n \geq 1}
    \frac{ q^{nj} }
         { \big( \frac{1-q^n}{1-q^{\phantom{n}}} \big)^{k} }
    {}\leq{}
    \begin{cases}
    \hfill 
    \dfrac{q^j}{1-q^j}
    \hfill
    &
    q < 1
    \\[1ex]
    \hfill
    \zeta(k)
    \vphantom{\dfrac{q^j}{1-q^j}}
    \hfill
    &
    q=1
    \\[1ex]
    \hfill
    \dfrac{ q^{j} } { 1-q^{j-k} }
    \hfill
    &
    q>1
    \end{cases}
\end{align}
For $q<1$, there is also a contribution from the non-meromorphic factor $q^{ \alpha_q(s) \alpha_q(t) }$ which appears in~\eqref{eq:CoonIP1} and~\eqref{eq:CoonIP2}. In total we find,
\begin{align}
\label{eq:CoonLE}
    \cA_{q}(s,t)
    &=
    \frac{1}{st} \,
    \exp
    \sum_{\ell_1, \ell_2 \geq 1}
    \bigg\{
    \Theta(1-q)
    \frac{ (1-q)^{\ell_1+\ell_2} }{ \ell_1 \ell_2 \ln q }
    -
    \sum_{j=1}^{\ell_{\min}}
    d_{j;\ell_1,\ell_2}
    \operatorname{Li}_{\ell_1+\ell_2}(q^j;q)
    \bigg\}
    \,
    s^{\ell_1}
    t^{\ell_2}
\\[1ex]
    &=
    \frac{1}{st}
    -
    \Big[
    \operatorname{Li}_2(q;q)
    - \Theta(1-q)
        \tfrac{(1-q)^2}{\ln q}
    \Big]
    -
    \Big[
    \operatorname{Li}_3(q;q)
    - \Theta(1-q)
        \tfrac{(1-q)^3}{2 \ln q}
    \Big]
    (s+t)
    + \cdots
\no
\end{align}
where $\ell_{\min} = \min(\ell_1,\ell_2)$ and $d_{j;\ell_1,\ell_2}$ is the following rational number,
\begin{align}
\label{eq:djll}
    d_{j;\ell_1,\ell_2}
    =
    \sum_{i=j}^{\ell_{\min}}
    \frac{ (\ell_1+\ell_2-i-1)! }
         { (\ell_1-i)! (\ell_2-i)! (i-j)! j! }
    \, (-)^{i-j}
\end{align}
The limit $q \to 1$ reproduces the low-energy expansion of the Veneziano amplitude~\eqref{eq:VenLE} since $\operatorname{Li}_{\ell_1+\ell_2}(1;1) = \zeta(\ell_1+\ell_2)$ and,
\begin{align}
    \sum_{j=1}^{\ell_{\min}}
    d_{j;\ell_1,\ell_2}
    =
    \frac{1}{\ell_1+\ell_2}
    \binom{\ell_1+\ell_2}{\ell_1}
\end{align}
for positive integers $\ell_1, \ell_2$.

\sm

The $q$-deformed polylogarithms, like the usual polylogarithms and the Riemann zeta-values, may be assigned a transcendental weight. If we assign weight $k$ to $\operatorname{Li}_{k}(q^j;q)$, then we must assign weight one to the factor $(1-q)$ since,
\begin{align}
    (1-q)
    \operatorname{Li}_{k}(q^j;q)
    =
    \operatorname{Li}_{k+1}(q^j;q)
    - \operatorname{Li}_{k+1}(q^{j+1};q)
\end{align}
Under these assignments, each side of this equation has weight $k+1$. If we assign weight~$-1$ to the Mandelstam variables as we did in the low-energy expansions of the Veneziano and Virasoro amplitudes, then each term in the low-energy expansion of the Coon amplitude with~${q \geq 1}$ has uniform transcendental weight two, just like the Veneziano amplitude~\eqref{eq:VenLE}.

\sm

For $q<1$, the transcendental structure is not as clear. In this case, the argument of the exponential (which should have transcendental weight zero) includes the terms,
\begin{align}
\label{eq:trans}
    \frac{1}{ \ell_1 \ell_2 \ln q } \,
    (1-q)^{\ell_1+\ell_2} \,
    s^{\ell_1} t^{\ell_2}
\end{align}
The factor $(1-q)^{\ell_1+\ell_2} \, s^{\ell_1} t^{\ell_2}$ has weight zero under our previous assignments, but it is customary to assign weight one to logarithms. After all, the logarithm is just the weight-one polylogarithm,
\begin{align}
    \operatorname{Li}_1(z) = -\ln(1-z)
\end{align}
so that~\eqref{eq:trans} naively has transcendental weight $-1$ rather than weight zero.

\sm

We are not, however, out of luck. We may write the reciprocal $1/\ell_1\ell_2$ in terms of finite harmonic sums,
\begin{align}
    H_1(k)
    =
    \sum_{n=1}^{k-1}
    \frac{1}{n}
    \quad
    \implies
    \quad
    \frac{1}{\ell_1\ell_2}
    =
    H_1(\ell_1\ell_2+1)
    - H_1(\ell_1\ell_2)
\end{align}
We then assign transcendental weight one to the finite harmonic sums so that~\eqref{eq:trans} has weight zero. This assignment is delicate. One should think of $H_1(k)$ not as its value for a single~$k$ (which is a rational number whose natural transcendental weight assignment is zero) but instead as a function of $k$ to be inserted into an infinite series in $k$. For instance,~$H_1(k)$ occurs in this manner in the double zeta-value $\zeta(\ell,1)$,
\begin{align}
    \zeta(\ell,1)
    =
    \sum_{n_1 > n_2 \geq 1}
    \frac{1}{n_1^\ell n_2^{\phantom{\ell}}}
    =
    \sum_{n \geq 2}
    \frac{H_1(n)}{n^\ell}
\end{align}
The standard weight assignments of $\zeta(\ell)$ and $\zeta(\ell,1)$ are $\ell$ and $\ell+1$, respectively, which justifies assigning weight one to the function $H_1(k)$. This assignment of non-zero transcendental weight to finite harmonic sums is familiar to the low-energy expansion of one-loop superstring amplitudes~\cite{DHoker:2019blr, DHoker:2021ous} and to loop amplitudes in $\cN = 4$ supersymmetric quantum field theory~\cite{Kotikov:2002ab, Beccaria:2009vt}.

\sm

Under these assignments, each term in the low-energy expansion of the Coon amplitude for all $q \geq 0$ has uniform transcendental weight two, in perfect analogy with the low-energy expansion of the Veneziano amplitude~\eqref{eq:VenLE}. For $q \geq 1$, the subtleties involving $\ln q$ and finite harmonic sums can be ignored.


\section{The Virasoro-Coon amplitude?}
\label{sec:CoonVir}

In this section, we shall attempt to construct a $q$-deformed Virasoro or Virasoro-Coon amplitude in analogy with our interpretation of the Coon amplitude as a $q$-deformed Veneziano amplitude. Specifically, we shall try to construct an amplitude~$\cA_{q\text{-Vir}}(s,t,u)$ with the following properties:
\begin{itemize}
\item $(s,t,u)$ crossing symmetry
\item simple poles in each channel only at the $q$-integers~$[n]_q$ with $n \geq 0$
\item polynomial residues on the massive poles
\item the field theory amplitude at $q=0$ and the Virasoro amplitude at $q=1$,
\begin{align}
    \cA_{q\text{-Vir}}(s,t,u)
    \quad
    & \xrightarrow[q \to 0]{}
    \quad
    \mask{\cA_{\text{Vir}}(s,t,u)}{-\frac{1}{stu}}
\no \\
    \cA_{q\text{-Vir}}(s,t,u)
    \quad
    & \xrightarrow[q \to 1]{}
    \quad
    \cA_{\text{Vir}}(s,t,u)
\end{align}
\item the low-energy expansion $-\frac{1}{stu} \, \big( 1+\cO(s,t,u) \big)$

\end{itemize}

We shall first consider the location of the poles. For $q \neq 1$, a convergent infinite product which contains our desired sequence of poles in each channel is,
\begin{align}
    \prod_{n \geq 0}
    \frac{1}{ \big( 1-\hat{q}^{\, n-\alpha_q(s)} \big)
              \big( 1-\hat{q}^{\, n-\alpha_q(t)} \big)
              \big( 1-\hat{q}^{\, n-\alpha_q(u)} \big) }
\end{align}
where $\hat{q} = \min(q,q^{-1})$. This infinite product, and thus $\cA_{q\text{-Vir}}$, is proportional to the following product of three $q$-gamma functions,
\begin{align}
\label{eq:qVir1}
    \Gamma_q \big( {-\alpha_q(s)} \big)
    \Gamma_q \big( {-\alpha_q(t)} \big)
    \Gamma_q \big( {-\alpha_q(u)} \big)
\end{align}
If we are to have a polynomial residue on each massive $s$-channel pole, then the infinite product of $t$-channel poles from $\Gamma_q \big( {-\alpha_q(t)} \big)$ must be canceled by an infinite product of zeroes. This cancellation can only be achieved by a function proportional to the ratio,
\begin{align}
\label{eq:qVir2}
   \frac{ \Gamma_q \big( {-\alpha_q(t)} \big) }
         { \Gamma_q \big( 1 - \alpha_q(s) - \alpha_q(t) \big) }
\end{align}
The Coon amplitude achieves polynomial residues through the same cancellation. The further requirement that $\cA_{q\text{-Vir}} = -\frac{1}{stu}$ at low-energy implies that all but $t^{-1}$ should be canceled from $\Gamma_q \big( {-\alpha_q(t)} \big)$ on the massless pole at $s = 0$ for all $q \geq 0$. We may satisfy this condition by multiplying~\eqref{eq:qVir2} by $q^{-\alpha_q(t)}$. Demanding $(s,t,u)$ symmetry, we find that $\cA_{q\text{-Vir}}$ must be proportional to,
\begin{align}
\label{eq:qVir3}
    q^{-\delta_q(s,t,u)} \,
    \frac{ \Gamma_q \big( {-\alpha_q(s)} \big) \,
           \Gamma_q \big( {-\alpha_q(t)} \big) \,
           \Gamma_q \big( {-\alpha_q(u)} \big) }
         { \Gamma_q \big( 1 - \alpha_q(t) - \alpha_q(u) \big) \,
           \Gamma_q \big( 1 - \alpha_q(u) - \alpha_q(s) \big) \,
           \Gamma_q \big( 1 - \alpha_q(s) - \alpha_q(t) \big) }
\end{align}
where $\delta_q(s,t,u) = \alpha_q(s) + \alpha_q(t) + \alpha_q(u)$. At $q=1$, this expression reproduces the Virasoro amplitude as desired.

\sm

Now on each massive $s$-channel pole, the factor $1/\Gamma_q \big( 1 - \alpha_q(t) - \alpha_q(u) \big)$ contributes an infinite product of zeroes in $t$, spoiling the polynomial residue. These zeroes do not appear if $q=0$ or $q=1$ because $\alpha_q(s) + \alpha_q(t) + \alpha_q(u) = 0$ when $q=0$ or $q=1$. For general~$q$, the infinite product of zeroes must be canceled by an infinite product of poles, and this cancellation can only be achieved by a function proportional to the ratio,
\begin{align}
    \frac{ \Gamma_q \big( \ell - \delta_q(s,t,u) \big) }
         { \Gamma_q \big( 1 - \alpha_q(t) - \alpha_q(u) \big) }
\end{align}
for some integer $\ell \geq 1$. While the factor $\Gamma_q \big( \ell - \delta_q(s,t,u) \big)$ cancels the infinite product of zeroes, it also introduces an infinite number of new poles, spoiling our initial assumption.

\sm

Despite this complication of additional poles, we shall proceed with $\ell = 1$. We now have the following ansatz for $\cA_{q\text{-Vir}}$,
\begin{align}
\label{eq:qVir4}
    q^{-\delta_q(s,t,u)} \,
    \frac{ \Gamma_q \big( {-\alpha_q(s)} \big) \,
           \Gamma_q \big( {-\alpha_q(t)} \big) \,
           \Gamma_q \big( {-\alpha_q(u)} \big) \,
           \Gamma_q \big( 1 - \delta_q(s,t,u) \big) }
         { \Gamma_q \big( 1 - \alpha_q(t) - \alpha_q(u) \big) \,
           \Gamma_q \big( 1 - \alpha_q(u) - \alpha_q(s) \big) \,
           \Gamma_q \big( 1 - \alpha_q(s) - \alpha_q(t) \big) }
\end{align}
which has the following convergent infinite product form for all $q \geq 0$,
\begin{align}
\label{eq:qVir5}
    -\frac{1}{stu \mathstrut}
    \prod_{n \geq 1}
    \frac{ \big( 1 - \hat{q}^{\, n-\alpha_q(t)-\alpha_q(u)} \big)
           \big( 1 - \hat{q}^{\, n-\alpha_q(u)-\alpha_q(s)} \big)
           \big( 1 - \hat{q}^{\, n-\alpha_q(s)-\alpha_q(t)} \big)
           \big( 1 - \hat{q}^{\, n} \big) }
         { \big( 1 - \hat{q}^{\, n-\alpha_q(s)} \big)
           \big( 1 - \hat{q}^{\, n-\alpha_q(t)} \big)
           \big( 1 - \hat{q}^{\, n-\alpha_q(u)} \big)
           \big( 1 - \hat{q}^{\, n-\delta_q(s,t,u)} \big) }
\end{align}
where again $\hat{q} = \min(q,q^{-1})$. This ansatz reproduces the field theory amplitude at $q=0$ and the Virasoro amplitude at $q=1$. The residues of this ansatz, however, are not polynomials. Near the massive pole at $s = [N]_q$,~\eqref{eq:qVir5} becomes,
\begin{align}
    -\frac{1}{[N]_q \mathstrut}
    \frac{1}{tu \mathstrut}
    \frac{1}{ \big( 1 - \hat{q}^{\, N-\alpha_q(s)} \big) }
    \prod_{n = 1}^{N-1}
    \frac{1}{ \big( 1 - \hat{q}^{\, n-N} \big) }
    \prod_{n = 1}^{N}
    \frac{ \big( 1 - \hat{q}^{\, n-N-\alpha_q(t)} \big)
           \big( 1 - \hat{q}^{\, n-N-\alpha_q(u)} \big) }
         { \big( 1 - \hat{q}^{\, n-N-\alpha_q(t)-\alpha_q(u)} \big) }
\end{align}
After some straightforward algebra, we see that the residue at $s = [N]_q$ is a non-polynomial rational function of $t$ unless $q=0$ or $q=1$.

\sm

We have thus failed to construct a $q$-deformed Virasoro amplitude under our stated assumptions. Therefore, we conclude that there is no amplitude with $(s,t,u)$ symmetry, simple poles at the $q$-integers, and polynomial residues. Only the field theory amplitude at~${q=0}$ (with no massive poles) and the Virasoro amplitude at~${q=1}$ (with poles at the integers) satisfy our constraints. It seems then that there is no $q$-deformed Virasoro or Virasoro-Coon amplitude.

\sm

In our companion work~\cite{NGLL:2022}, we revisit this question by analyzing so-called generalized Virasoro amplitudes, defined by a generalization of the infinite product representation of the Virasoro amplitude~\eqref{eq:VirIP}. In this analysis, we assume $(s,t,u)$ symmetry and demand physical residues on an a priori unspecified sequence of poles~$ \lambda_n $. In other words, we do not assume a given mass spectrum as we have done in our search for a $q$-deformed Virasoro amplitude here. We find that the poles~$\lambda_n $ must satisfy an over-determined set of non-linear recursion relations. We then numerically demonstrate that the only consistent solution to these recursion relations is the string spectrum with $\lambda_n = n$.

\newpage

\appendix

\section{Gamma function}
\label{sec:Gamma}

In this appendix, we shall collect some well-known properties of the gamma function $\Gamma(z)$. The gamma function is a meromorphic function with poles at the non-positive integers, defined by the following integral,
\begin{align}
    \Gamma(z)
    =
    \int_0^\infty dx \,
    x^{z-1} e^{-x}
\end{align}
The gamma function obeys the functional equation,
\begin{align}
\label{eq:GammaFunc}
    \Gamma(z+1)
    =
    z \, \Gamma(z)
\end{align}
and the reflection formula,
\begin{align}
\label{eq:GammaRefl}
    \Gamma(z) \,
    \Gamma(1-z)
    =
    \frac{\pi}{\sin(\pi z)}
\end{align}
A useful infinite product representation of the gamma function is,
\begin{align}
\label{eq:GammaIP}
    \Gamma(z)
    =
    \frac{1}{z}
    \prod_{n=1}^\infty
    \frac{ \big( 1+\frac{1}{n} \big)^z }
         { \phantom{\big(} 1+\frac{z}{n} \phantom{\big)^z} }
\end{align}
A Taylor expansion for $\Gamma(1+z)$ with $|z|<1$ is given by,
\begin{align}
\label{eq:GammaTay}
    \ln \Gamma(1+z)
    =
    -\gamma_E \, z
    +\sum_{k=2}^\infty \frac{\zeta(k)}{k} \, (-z)^k
\end{align}
where $\gamma_E$ is the Euler-Mascheroni constant and $\zeta(k) = \sum_{n=1}^\infty n^{-k}$ are Riemann zeta-values. The asymptotic behavior of the gamma function is given by Stirling's formula,
\begin{align}
\label{eq:Stirling}
    \ln \Gamma(z)
    \widesim[2]{|z| \to \infty}
    (z-\half) \ln z
    - z
    + \half \ln(2\pi)
    + \cO \big( z^{-1} \big)
\end{align}
which is valid for large $|z| \gg 1$ with phase $|\arg(z)| < \pi$.

\newpage

\section{Gegenbauer polynomials}
\label{sec:Geg}

In this appendix, we shall review some properties of the Gegenbauer polynomials~\cite{Vilenkin1968, gradshteyn2007}. The Gegenbauer polynomials may be defined by a generating function,
\begin{align}
    \frac{1}{ (1-2xt+t^2)^{\lambda} }
    =
    \sum_{j=0}^\infty
    C^{(\lambda)}_j(x) \,
    t^j
\end{align}
or in terms of the hypergeometric function,
\begin{align}
    C^{(\lambda)}_j(x)
    &=
    \tfrac{ \Gamma(j+2\lambda) \vphantom{\frac{1}{2}} }
          { \Gamma(j+1) \Gamma(2\lambda) \vphantom{\frac{1}{2}} }
    \,
    {}_2 F_1
    \big( {-j} , \, j+2\lambda ; \, \lambda+\half ; \, \half(1-x) \big)
\end{align}
The first few Gegenbauer polynomials are,
\begin{align}
    C^{(\lambda)}_0(x)
    &=
    1
\no \\
    C^{(\lambda)}_1(x)
    &=
    2 \lambda \, x
\no \\
    C^{(\lambda)}_2(x)
    &=
    -\lambda + 2\lambda(1+\lambda) \, x^2
\no \\
    C^{(\lambda)}_3(x)
    &=
    - 2\lambda(1+\lambda) \, x
    + \tfrac{4}{3}\lambda(1+\lambda)(2+\lambda) \, x^3
\end{align}
In $d \geq 3$ spacetime dimensions, the polynomials $\smash{C_j^{(\frac{d-3}{2})}(\cos\theta)}$ diagonalize the Lorentz group Casimir operator. In $d=3$ we must omit the normalization factor $\smash{\tfrac{\Gamma(j+2\lambda)}{\Gamma(j+1)\Gamma(2\lambda)}}$ which vanishes. The case $d=4$ reduces to the familiar Legendre polynomials. The Gegenbauer polynomials obey an orthogonality relationship,
\begin{align}
\label{eq:GegI1}
    \int_{-1}^1 dx \, (1-x^2)^{\lambda-\frac{1}{2}} \,
    C^{(\lambda)}_j(x) \,
    C^{(\lambda)}_{\ell}(x)
    &=
    \begin{cases}
    \mask{
    \frac{ \Gamma(\lambda+\frac{1}{2}) \Gamma(j+2\lambda) \Gamma(\ell+1)
           \Gamma(\frac{\ell-j+1}{2}) }
         { 2^j \, \Gamma(2\lambda) \Gamma(j+1) \Gamma(\ell-j+1)
           \Gamma(\frac{\ell+j}{2}+\lambda+1) }
    }{
    \frac{ \pi \mathstrut }
         { 2^{2\lambda-1} (j+\lambda) \vphantom{\frac{1}{2}} }
    \frac{ \Gamma(j+2\lambda) }
         { \Gamma(j+1) \Gamma(\lambda)^2 \vphantom{\frac{1}{2}} } }
    &
    j=\ell
    \\
    \hfill 0 \hfill
    &
    j \neq \ell
    \end{cases}
\intertext{and the following integration identity,}
\label{eq:GegI2}
    \int_{-1}^1 dx \, (1-x^2)^{\lambda-\frac{1}{2}} \,
    C^{(\lambda)}_j(x) \,
    x^\ell
    &=
    \begin{cases}
    \frac{ \Gamma(\lambda+\frac{1}{2})
           \Gamma(j+2\lambda)
           \Gamma(\ell+1)
           \Gamma(\frac{\ell-j+1}{2}) }
         { 2^j \, 
           \Gamma(2\lambda)
           \Gamma(j+1)
           \Gamma(\ell-j+1)
           \Gamma(\frac{\ell+j}{2}+\lambda+1) }
    &
    j+\ell \text{ even} 
    \\
    \hfill 0 \hfill
    &
    j+\ell \text{ odd} 
    \end{cases}
\end{align}
for integers $j,\ell \geq 0$. These two integrals may be used to write the residues of any tree-level four-point amplitude in terms of Gegenbauer polynomials. The product of two Gegenbauer polynomials may be expanded as,
\begin{align}
    C^{(\lambda)}_{j_1}(x) \,
    C^{(\lambda)}_{j_2}(x)
    &=
    \sum_{\ell = |j_1-j_2|}^{j_1+j_2}
    c_{j_1,j_2;\ell}^{(\lambda)} \,
    C^{(\lambda)}_{\ell \mathstrut}(x)
\end{align}
for integers $j_1, j_2 \geq 0$, where,
\begin{align}
    c_{j_1,j_2;\ell}^{(\lambda)}
    &=
    \begin{cases}
    \frac{ (\ell+\lambda) \, \Gamma(\ell+1)
           \Gamma(g+2\lambda) \vphantom{\frac{1}{2}} }
         { \Gamma(\lambda)^2 \Gamma(\ell+2\lambda)
           \Gamma(g+\lambda+1) \vphantom{\frac{1}{2}} }
    \frac{ \Gamma(g-\ell+\lambda) \vphantom{\frac{1}{2}} }
         { \Gamma(g-\ell+1) \vphantom{\frac{1}{2}} }
    \frac{ \Gamma(g-j_1+\lambda) \vphantom{\frac{1}{2}} }
         { \Gamma(g-j_1+1) \vphantom{\frac{1}{2}} }
    \frac{ \Gamma(g-j_2+\lambda) \vphantom{\frac{1}{2}} }
         { \Gamma(g-j_2+1) \vphantom{\frac{1}{2}} }
    &
    j_1+j_2+\ell \text{ even}
    \\
    \hfill 0 \hfill
    &
    j_1+j_2+\ell \text{ odd} 
    \end{cases}
\end{align}
with $g=\half(j_1+j_2+\ell)$. For $d \geq 4$, the coefficients $\smash{c_{j_1,j_2;\ell}^{(\frac{d-3}{2})}} \geq 0$ are non-negative.

\newpage

\section{Deriving the Coon amplitude low-energy expansion}
\label{sec:CoonLE}

In this appendix, we shall derive the low-energy expansion~\eqref{eq:CoonLE} of the Coon amplitude for all $q \geq 0$. Our starting point is~\eqref{eq:CoonIP2}. The low-energy expansion of the factor $\smash{q^{ \alpha_q(s) \alpha_q(t)}}$ may be computed using the Taylor expansion for $\ln(1-z)$,
\begin{align}
    q^{ \frac{ \ln ( 1+(q-1)s ) }{ \ln q }
        \frac{ \ln ( 1+(q-1)t ) }{ \ln q } }
    =
    \exp
    \sum_{\ell_1, \ell_2 \geq 1}
    \frac{ (1-q)^{\ell_1+\ell_2} }{ \ell_1 \ell_2 \ln q }
    \,
    s^{\ell_1}
    t^{\ell_2}
\end{align}
The low-energy expansion of the infinite product is similarly given by,
\begin{align}
    & \quad
    \prod_{n \geq 1}
    \frac{ \big( \frac{1-q^n}{1-q^{\phantom{n}}} \big)^2
            - \big( \frac{1-q^n}{1-q^{\phantom{n}}} \big) (s+t)
            + (1-q^n) st }
        { \big( s - \frac{1-q^n}{1-q^{\phantom{n}}} \big)
          \big( t - \frac{1-q^n}{1-q^{\phantom{n}}} \big) }
\no \\[1ex]
    &=
    \exp \,
    \sum_{n \geq 1}
    \sum_{k \geq 1} \,
    \frac{1}{ k \, \big( \frac{1-q^n}{1-q^{\phantom{n}}} \big)^k }
    \Big[ s^k + t^k - \big( s+t+(q-1)st \big)^k \Big]
\end{align}
At this point we cannot interchange the sums over $n$ and $k$ and perform the sum over~$n$ because the resultant $q$-deformed polylogarithms $\operatorname{Li}_k(1;q)$ diverge for $q<1$. Instead, we expand the summand using the multinomial theorem and collect powers of $s$ and $t$ to find,
\begin{align}
    \exp \,
    \sum_{ n \geq 1 } \,
    \sum_{\ell_1, \ell_2 \geq 1} \,
    \sum_{ i=0 }^{ \ell_{\min} } \,
    \frac{ (\ell_1+\ell_2-i-1)! }
         { (\ell_1-i)! (\ell_2-i)! i! }
    \,
    \frac{ (-)(q^n-1)^i }
         { \big( \frac{1-q^n}{1-q^{\phantom{n}}} \big)^{\ell_1+\ell_2} }
    \,
    s^{\ell_1} t^{\ell_2}
\end{align}
where $\ell_{\min} = \min(\ell_1,\ell_2)$. We now expand the factor $(q^n-1)^i$ and find,
\begin{align}
    \exp \,
    \sum_{ n \geq 1 } \,
    \sum_{\ell_1, \ell_2 \geq 1} \,
    \sum_{j=1}^{ \ell_{\min} } \,
    d_{j;\ell_1,\ell_2}
    \,
    \frac{(-) \, q^{nj}}
         { \big( \frac{1-q^n}{1-q^{\phantom{n}}} \big)^{\ell_1+\ell_2} }
    \,
    s^{\ell_1} t^{\ell_2}
\end{align}
with the rational numbers $d_{j;\ell_1,\ell_2}$ defined in~\eqref{eq:djll}. The $j=0$ terms vanish because,
\begin{align}
    d_{0;\ell_1,\ell_2}
    =
    \sum_{i=0}^{\ell_{\min}}
    \frac{ (\ell_1+\ell_2-i-1)! }
         { (\ell_1-i)! (\ell_2-i)! i! }
    \, (-)^{i}
    =
    0
\end{align}
We may now interchange the order of the infinite sums and perform the sum over $n$ because the resultant $q$-deformed polylogarithms $\operatorname{Li}_k(q^j;q)$ are absolutely convergent for all $q \geq 0$. Combining our results, we arrive at~\eqref{eq:CoonLE}.

\newpage

\bibliography{InfProdAmp}

\end{document}

%% file: preambledefs.tex
\newcommand{\cA}{\mathcal{A}}

\newcommand{\cN}{\mathcal{N}}
\newcommand{\cO}{\mathcal{O}}

\renewcommand{\[}{\left[}

\DeclareMathOperator*{\Res}{Res}

\newcommand{\widesim}[2][1.5]{
  \mathrel{\overset{#2}{\scalebox{#1}[1]{$\sim$}}}
}

\newcommand{\half}{\tfrac{1}{2}}

\newcommand{\negphantom}[1]{
    \ifmmode\settowidth{\dimen0}{$#1$}
    \else\settowidth{\dimen0}{#1}
    \fi
    \hspace*{-\dimen0}}
    
\makeatletter
\newcommand{\mask}[2]{{\mathpalette\mask@{{#1}{#2}}}}
\newcommand{\mask@}[2]{\mask@@{#1}#2}
\newcommand{\mask@@}[3]{%
  \settowidth{\dimen@}{$\m@th#1#2$}%
  \makebox[\dimen@]{$\m@th#1#3$}%
}
\makeatother
    
\newcommand{\sm}{\smallskip}
\newcommand{\no}{\nonumber}
